\documentclass[aps,prd]{revtex4}

\usepackage{amsfonts,amsmath}
\usepackage{graphicx}

\begin{document}

\title{Constraints on a Cardassian model from SNIa data -- revisited}

\author{W{\l}odzimierz God{\l}owski}
\email{godlows@oa.uj.edu.pl}

\author{Marek Szyd{\l}owski}
\email{uoszydlo@cyf-kr.edu.pl}

\affiliation{Astronomical Observatory, Jagiellonian University,
Orla 171, 30-244 Krak{\'o}w, Poland}

\author{Adam Krawiec}
\email{uukrawie@cyf-kr.edu.pl}
\affiliation{Institute of Public Affairs, Jagiellonian University,
Rynek G{\l}{\'o}wny 8, 31-042 Krak{\'o}w, Poland}

\begin{abstract}
We discuss some observational constraints, resulting from SN Ia observations, 
imposed on the behavior of the original flat Cardassian model, and its 
extension with the curvature term included. We test the models using the 
Perlmutter SN Ia data as well as the new Knop and Tonry samples. We estimate 
the Cardassian model parameters using the best-fitting procedure and the 
likelihood method. In the fitting procedure we use density variables for 
matter, Cardassian fluid and curvature, and include the errors in redshift 
measurement. For the Perlmutter sample in the non-flat Cardassian model we 
obtain the high or normal density universe ($\Omega_{\mathrm{m},0} \ge 0.3$), 
while for the flat Cardassian model we have the high density universe. 
For sample A in the high density universe we also find the negative values of 
estimates of $n$ which can be interpreted as the phantom fluid effect. For the 
likelihood method we get that a nearly flat universe is preferred. We show 
that, if we assume that the matter density is $0.3$, then $n \approx 0$ 
in the flat Cardassian model, which corresponds to the Perlmutter model with 
the cosmological constant. Testing with the Knop and Tonry SN Ia samples 
show no significant differences in results.
\end{abstract}

\maketitle

\section{Introduction}

\citet{Freese02} have recently proposed an alternative to the cosmological
constant model explaining the currently accelerating Universe.
In this model the standard Friedmann-Robertson-Walker (FRW) equation is
modified by the presence of an additional term $\rho^n$, namely
\begin{equation}
\label{eq:1}
H^2 = \frac{\rho}{3} + B\rho^n,
\end{equation}
where $H \equiv \dot{a}/a$ is the Hubble parameter, $a$ is the scale factor,
$\rho$ is the energy density of matter and radiation, and $B$ is a positive 
constant. For simplicity it is assumed that, the density parameter for 
radiation matter $\Omega_{\mathrm{r},0}=0$. This proposal seems to be 
attractive because the expansion of the universe is accelerated automatically 
due to the presence of the additional term (if we put $B=0$ then the standard 
FRW equation is recovered).

Because the Cardassian model offers an alternative to the cosmological constant
model, the agreement of this model with available observations of type Ia 
supernovae (SN Ia) was immediately verified \citep{Zhu03,Sen03}. Some 
interesting results on observational constraints in the generalized Cardassian 
model were also obtained from the statistical analysis of gravitational lensed 
quasars \citep{Dev03}. In our analysis we use three SN Ia data sets compiled 
by \citet{Perlmutter99}, \citet{Knop03} and \citet{Tonry03} to test both the 
original Cardassian model and its extension with a curvature term. Our 
results are inconsistent with \citet{Zhu03}'s prediction of the low density 
Universe. The noteworthy results from the joint analysis of SN Ia data and 
CMBR were obtained by \citet{Sen03} who also confirmed the prediction of 
low density Universe, but their analysis was restricted to the case of 
$n > 0$. This case was also examined by \citet{Frith03} who used the Tonry 
SN Ia data. However, such a choice of an interval for this parameter is not 
physically justified and both negative and positive values are admissible.

\section{Basic equation}

We add a curvature term to the original Cardassian model.
Then equation~(\ref{eq:1}) assumes the following form
\begin{equation}
\label{eq:2}
H^2= \frac{\rho}{3}+B\rho^n+\frac{k}{a^2}.
\end{equation}
It is useful to rewrite equation~(\ref{eq:2}) in dimensionless variables
$\Omega_{i,0}$, $i=(\mathrm{m},\mathrm{Card},k)$,
\begin{equation}
\label{eq:3}
\frac{H^2}{H_0^2}=\Omega_{\mathrm{m},0}\left(\frac{a}{a_0}\right)^{-3}
+\Omega_{\mathrm{Card},0}\left(\frac{a}{a_0}\right)^{-3n}
+\Omega_{k,0}\left(\frac{a}{a_0}\right)^{-2},
\end{equation}
where energy density for the dust is assumed ($p=0$) (and hence
$\rho \propto a^{-3}$ from a conservation equation), $\Omega_{\mathrm{m},0}$
is the matter density parameter; $\Omega_{\mathrm{Card},0}=3B\rho^n/3H_0^2$
is the density parameter for the fictitious noninteracting fluid which
mimics term $\rho^n$ and yields additional density in the model.
The subscript zero denotes a present value of model parameters. Then
\begin{equation}
\label{eq:4}
\dot{\rho}=-3H(\rho+p)
\end{equation}
For $a=a_0$ (the present value of scale factor) we obtain the following
constraint
\begin{equation}
\label{eq:5}
\Omega_{\mathrm{m},0} + \Omega_{\mathrm{Card},0} + \Omega_{k,0} = 1.
\end{equation}
For the fictitious Cardassian fluid we obtain $p=(n-1)\rho$ if
$\rho \propto a^{-3}$. In this interpretation the negative value of 
$n$ is equivalent to the existence of phantom matter with supernegative 
presssure. 

Let us note that in the analogous equation describing FRW dynamics with
dust and additional fictitious fluid with the equation of state $p=(n-1)\rho$
presented in \citet{Sen03} only case $n>0$ was considered.

\section{Magnitude-redshift relation}

It is well known that various cosmic distance measures, for example the
luminosity distance, depend sensitively both on the spatial geometry
(curvature) and dynamics. Therefore, luminosity depends on the present
densities of the different components of matter content and its equation
of state. For this reason, the magnitude-redshift relation for distant
objects is proposed as a potential test for cosmological models and plays
an important role in determining cosmological parameters.

Let us consider an observer located at $r=0$ at the moment $t=t_0$ who
receives light emitted at $t=t_1$ from a source of absolute luminosity $L$
located at the radial distance $r_1$. It is known that the cosmological
redshift $z$ of the source is related to $t_1$ and $t_0$ by the relation
$1+z=a(t_0)/a(t_1)$. If the apparent luminosity $l$ of the source measured by
the observer is  defined by
\begin{equation}
\label{eq:6}
l = \frac{L}{4\pi d_L^2}
\end{equation}
then the luminosity distance $d_L$ of the source is
\begin{equation}
\label{eq:7}
d_L = (1+z)a_0 r_1.
\end{equation}

For historical reasons, the observed and absolute luminosities are defined
in terms of K-corrected observed magnitudes and absolute magnitudes $m$ and
$M$, respectively ($l=10^{-2m/5}\cdot 2.52 \cdot 10^{-5}\,\mathrm{erg}\,
\mathrm{cm}^{-2}\,\mathrm{s}^{-2}$, $L=10^{-2M/5}\cdot 3.02 \cdot
10^{35}\,\mathrm{erg}\,\mathrm{s}^{-2}$) \citep{Weinberg72}.
When written in terms of $m$ and $M$, equation~(\ref{eq:6}) yields
\begin{equation}
\label{eq:8}
m(z,\mathcal{M},\Omega_{\mathrm{m},0},\Omega_{\mathrm{Card},0},\Omega_{k,0})
= \mathcal{M} + 5\log_{10}[\mathcal{D}_L(z,\Omega_{\mathrm{m},0},
\Omega_{\mathrm{Card},0},\Omega_{k,0})]
\end{equation}
where
\begin{equation}
\label{eq:9}
\mathcal{M}=M-5\log_{10}H_0+25
\end{equation}
and
\begin{equation}
\label{eq:10}
\mathcal{D}_L(z,\Omega_{\mathrm{m},0},\Omega_{\mathrm{Card},0},\Omega_{k,0})
\equiv H_0 d_L(z,\Omega_{\mathrm{m},0},\Omega_{\mathrm{Card},0},\Omega_{k,0},H_0)
\end{equation}
is the dimensionless luminosity distance while $d_L$ is in Mpc.

The standard analysis yields the following relationship for the
dimensionless luminosity distance
\begin{multline}
\mathcal{D}_L((z,\Omega_{\mathrm{m},0},\Omega_{\mathrm{Card},0},\Omega_{k,0})
= \frac{(1+z)}{\sqrt{\mathcal{K}}}\xi\left(\sqrt{\mathcal{K}}\int\limits_0^z
[(1-\Omega_{\mathrm{m},0}-\Omega_{\mathrm{Card},0})(1+z')^2 \right. \\
\label{eq:11}
\left. \phantom{\int\limits_0^z}
+ \Omega_{\mathrm{m},0}(1+z')^3 + \Omega_{\mathrm{Card},0}(1+z')^{3n}
]^{-1/2}dz' \right),
\end{multline}
where
\begin{align}
\xi(x) &= \sin x& \mathrm{with}& & \mathcal{K}&=-\Omega_{k,0}&
\mathrm{when}& &\Omega_{k,0}&<0 \nonumber \\
\xi(x) &= x& \mathrm{with}& & \mathcal{K}&=1&
\mathrm{when}& &\Omega_{k,0}&=0
\label{eq:12} \\
\xi(x) &= \sinh x& \mathrm{with}& & \mathcal{K}&=\Omega_{k,0}&
\mathrm{when}& &\Omega_{k,0}&>0 \nonumber
\end{align}
and the density parameter for hypothetical curvature fluid is
\[
\Omega_{k,0} = - \frac{k}{\dot{a}_0^2}.
\]
Thus, for given $\mathcal{M}$, $\Omega_{\mathrm{m},0}$,
$\Omega_{\mathrm{Card},0}$, $\Omega_{k,0}$, equations (\ref{eq:8}) and
(\ref{eq:11}) give the predicted value of $m(z)$ at a given $z$.

The goodness of fit is characterized by the parameter
\begin{equation}
\label{eq:13}
\chi^{2}=\sum_{i} \frac{|\mu_{0,i}^{0}-\mu_{0,i}^{t}|}{\sigma_{\mu 0,i}^{2}
+ \sigma_{\mu z,i}^{2}}
\end{equation}
where $\mu_{0,i}^{0}$ is the measured value, $\mu_{0,i}^{t}$
is the value calculated in the model described above, $\sigma_{\mu 0,i}^{2}$
is the measurement error, $\sigma_{\mu z,i}^{2}$ is the dispersion in the
distance modulus due to peculiar velocities of galaxies.

We assume that supernovae measurements come with uncorrelated Gaussian errors 
and in this case the likelihood function $\mathcal{L}$ can be determined from 
a chi-squared statistic $\mathcal{L} \propto \exp{(-\chi^{2}/2)}$ 
\citep{Perlmutter99,Riess98}.

\section{Statistical analysis with the Perlmutter sample}

We have decided to test our model using the Perlmutter samples of supernovae
\citep{Perlmutter99}. We estimate the value of $\mathcal{M}$
(equation~(\ref{eq:9})) separately for the full sample of 60 supernovae
(sample A) and for the sample of 54 supernovae (sample C --- in that sample
we exclude 2 supernovae as outliers and 2 as likely reddened ones from
the sample of 42 high redshift supernovae and 2 outliers from the sample
of 18 low redshift supernovae). We obtain the value of $\mathcal{M}=-3.39$
for sample A, and $\mathcal{M}=-3.43$ for sample C. To test the model we
calculate the best fit with minimum $\chi^2$ as well as estimate the model
parameters using the likelihood method \citep{Riess98}. For both statistical
methods we take the parameter $n$ in the interval $[-3.33, 2]$,
$\Omega_{\mathrm{m},0}$ in the interval $[0, 1]$, $\Omega_{k,0}$ in the
interval $[-1, 1]$, while an interval for $\Omega_{\mathrm{Card},0}$ is
obtained from equation~(\ref{eq:5}).

First, we estimate best fitting parameters for the non-flat Cardassian model
and obtain $\Omega_{k,0}=-1.00$, $\Omega_{\mathrm{m},0}=0.83$, $n=-0.67$,
$\Omega_{\mathrm{Card},0}=1.17$ for sample A, and
$\Omega_{k,0}=-1.00$, $\Omega_{\mathrm{m},0}=0.60$, $n=0.50$,
$\Omega_{\mathrm{Card},0}=1.40$ for sample C. This model is characterized
with strong negative curvature, that is similar to the Perlmutter results.
Let us note that such a high value of $\Omega_{k,0}$ is in contradiction
with CMB observations, which prefer the flat universe. When we assume that
$\Omega_{k,0}=0$ then we obtain the best-fitted flat Cardassian model with
$\Omega_{\mathrm{m},0}=0.52$, $n=-0.93$, $\Omega_{\mathrm{Card},0}=0.48$ for
sample A, and $\Omega_{\mathrm{m},0}=0.46$, $n=-0.60$,
$\Omega_{\mathrm{Card},0}=0.54$ for sample C. The detailed values are in
Tables~\ref{tab:1} and~\ref{tab:2}.

\begin{table}
\noindent
\caption{Results of the statistical analysis of the Cardassian model
obtained both for Perlmutter sample A and C from the best fit with minimum
$\chi^2$ (denoted with BF) and from the likelihood method (denoted with L).
The same analysis was repeated with fixed $\Omega_{\mathrm{m},0}=0.3$.}
\label{tab:1}
\begin{tabular}{@{}p{1.5cm}rrrrrrr}
\hline \hline
sample & $\Omega_{\mathrm{Card}}$ & $\Omega_{k,0}$ &
$\Omega_{\mathrm{m},0}$ & $n$ & $\mathcal{M}$ & $\chi^2$& method \\
\hline
  A   &  1.17 &-1.00 & 0.83 &-0.67 &-3.39 & 94.8 &  BF  \\
      &  0.37 &-0.12 & 0.50 &-0.13 &-3.39 & ---  &  L   \\
      &  1.65 &-0.95 & 0.30 & 0.37 &-3.39 & 95.7 &  BF  \\
      &  0.27 & 0.45 & 0.30 & 0.33 &-3.39 & ---  &  L   \\
\hline
  C   &  1.40 &-1.00 & 0.60 & 0.50 &-3.43 & 52.8 &  BF  \\
      &  0.33 & 0.17 & 0.29 & 0.10 &-3.43 & ---  &  L   \\
      &  1.70 &-1.00 & 0.30 & 0.36 &-3.43 & 53.2 &  BF  \\
      &  0.29 & 0.41 & 0.30 & 0.30 &-3.43 & ---  &  L   \\
\end{tabular}
\end{table}

\begin{table}
\noindent
\caption{Results of the statistical analysis of the Cardassian flat model
for Perlmutter sample A and sample C obtained both from the best fit with
minimum $\chi^2$ (denoted with BF) and from the likelihood method (denoted 
with L). In the case in which we marginalize over $\mathcal{M}$ we denote 
it with $\mathcal{M}$. The same analysis was repeated with fixed
$\Omega_{\mathrm{m},0}=0.3$.}
\label{tab:2}
\begin{tabular}{@{}p{1.5cm}rrrrrr}
\hline \hline
sample & $\Omega_{\mathrm{Card},0}$ & $\Omega_{\mathrm{m},0}$ &
$n$ & $\mathcal{M}$ & $\chi^2$& method \\
\hline
  A   &  0.48 & 0.52 & -0.93 & -3.39 & 95.3 & BF               \\
      &  0.40 & 0.60 & -0.53 & -3.39 & ---  & L                \\
      &  0.46 & 0.54 & -1.43 & -3.44 & 95.0 & $\mathcal{M}$, BF\\
      &  0.46 & 0.54 & -1.20 & -3.41 & ---  & $\mathcal{M}$, L \\
      &  0.70 & 0.30 &  0.03 & -3.39 & 96.1 & BF               \\
      &  0.70 & 0.30 &  0.03 & -3.39 & ---  & L                \\
      &  0.70 & 0.30 &  0.03 & -3.39 & 96.1 & $\mathcal{M}$, BF\\
      &  0.70 & 0.30 &  0.03 & -3.39 & ---  & $\mathcal{M}$, L \\
\hline
  C   &  0.54 & 0.46 & -0.60 & -3.43 & 95.3 & BF               \\
      &  0.42 & 0.58 &  0.10 & -3.43 & ---  & L                \\
      &  0.52 & 0.48 & -0.96 & -3.46 & 95.0 & $\mathcal{M}$, BF\\
      &  0.48 & 0.52 &  0.17 & -3.45 & ---  & $\mathcal{M}$, L \\
      &  0.7  & 0.30 & -0.03 & -3.43 & 53.3 & BF               \\
      &  0.7  & 0.30 & -0.03 & -3.43 & ---  & L                \\
      &  0.7  & 0.30 & -0.03 & -3.43 & 53.3 & $\mathcal{M}$, BF\\
      &  0.7  & 0.30 & -0.03 & -3.43 & ---  & $\mathcal{M}$, L \\
\end{tabular}
\end{table}

\begin{figure}
\includegraphics[width=0.8\textwidth]{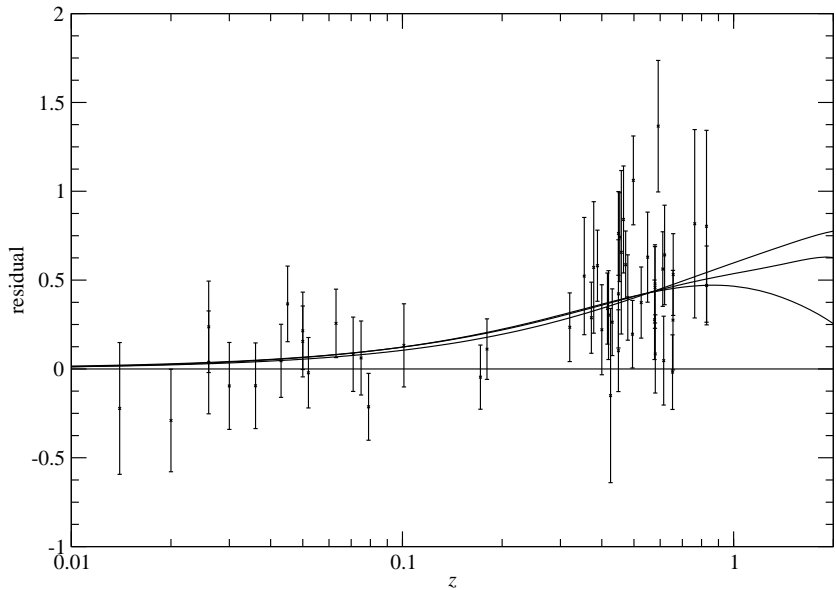}
\caption{Residuals (in mag) between the Einstein-de Sitter model and four
cases: the Einstein-de Sitter itself (zero line), the Perlmutter flat model
(upper curve), the best-fitted flat Cardassian model (upper-middle curve),
$\Omega_{k,0}=0$, $\Omega_{\mathrm{m},0}=0.46$, $n=-0.60$,
$\Omega_{\mathrm{Card},0}=0.54$ and the best-fitted (non-flat) Cardassian
model (lower-middle curve)
$\Omega_{k,0}=-1.00$, $\Omega_{\mathrm{m},0}=0.60$, $n=0.50$,
$\Omega_{\mathrm{Card},0}=1.40$.}
\label{fig:1}
\end{figure}

For sample C we draw the best-fitted, non-flat (upper-middle curve), flat
(lower-middle curve) Cardassian models as well as the best-fitted Perlmutter
model with $\Omega_{\mathrm{m},0}=0.28$, $\Omega_{\Lambda,0}=0.72$ (upper
curve) against to the flat Einstein-de Sitter model (zero line) on the Hubble
diagram (Fig.~\ref{fig:1}). One can observe that the difference between the
best-fitted Cardassian model and the Einstein-de Sitter model with
$\Omega_{\Lambda,0}=0$ assumes the largest value for $z \sim 0.9$ and
significantly decreases for higher redshifts. While the best-fitted flat
Cardassian model and Perlmutter one have increasing differences to the flat
Einstein-de Sitter model for higher redshifts, the differences between the
best-fitted flat Cardassian model and Perlmutter model increase for higher
redshifts. It gives us a possibility to discriminate between the Perlmutter
model and Cardassian models if $z \simeq 1$ supernovae data will be available.
It is very important because for the present data the Cardassian models are
only marginally better than the Perlmutter model.

However, knowing the best-fit values alone has no enough scientific relevance,
if confidence levels for parameter intervals are not presented, too.
Therefore, we carry out the model parameters estimation using the minimization
procedure, based on the likelihood method. On the confidence level $68.3\%$ we
obtain parameter values for samples A and C (Table~\ref{tab:3}).

\begin{table}
\caption{Non-flat Cardassian model parameter values obtained from the 
minimization procedure carried out on the Perlmutter samples.}
\label{tab:3}
\begin{tabular}{@{}p{1.5cm}rrrr}
\hline \hline
sample & $\Omega_{k,0}$ & $\Omega_{\mathrm{Card},0}$ &
$\Omega_{\mathrm{m},0}$ & $n$ \\
\hline
A &  $-0.12^{+0.54}_{-0.50}$ & $0.37^{+0.38}_{-0.20}$
  & $0.50^{+0.28}_{-0.32}$ & $-0.13^{+0.56}_{-1.37}$ \\
C & $0.17^{+0.44}_{-0.66}$ & $0.33^{+0.43}_{-0.17}$
  & $0.29^{+0.28}_{-0.28}$ & $0.10^{+0.36}_{-1.46}$
\end{tabular}
\end{table}

\begin{figure}
\includegraphics[width=0.6\textwidth]{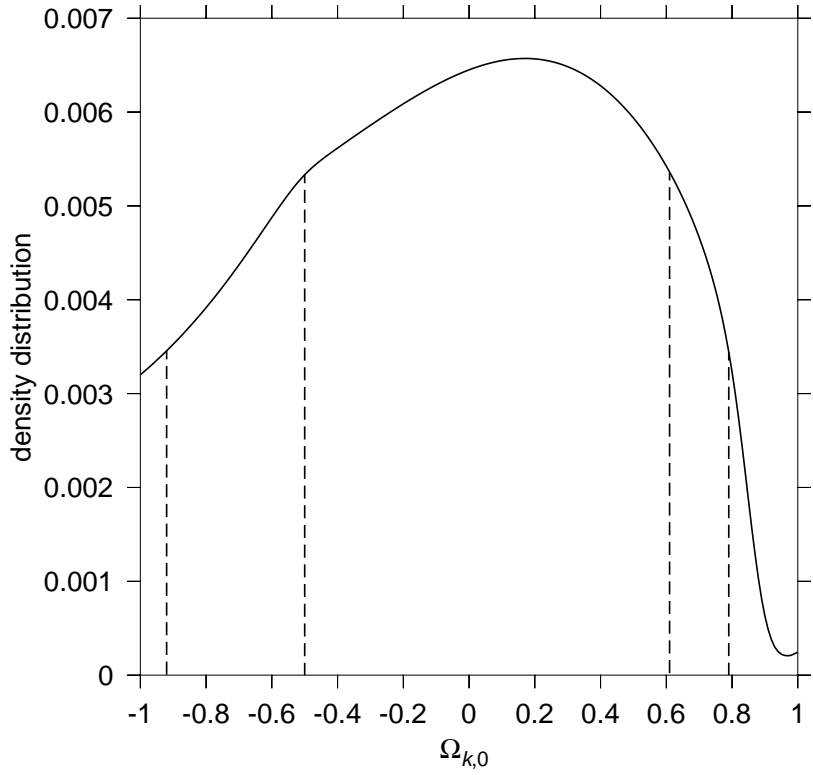}
\caption{The density distribution for $\Omega_{k,0}$ in the Cardassian
model (sample C). We obtain that $\Omega_{k,0}=0.17^{+0.46}_{-0.66}$
on the confidence level $68.3\%$ (the inner dash lines).
Both positive and negative values of $\Omega_{k,0}$ are formally possible.}
\label{fig:2}
\end{figure}

\begin{figure}
\includegraphics[width=0.6\textwidth]{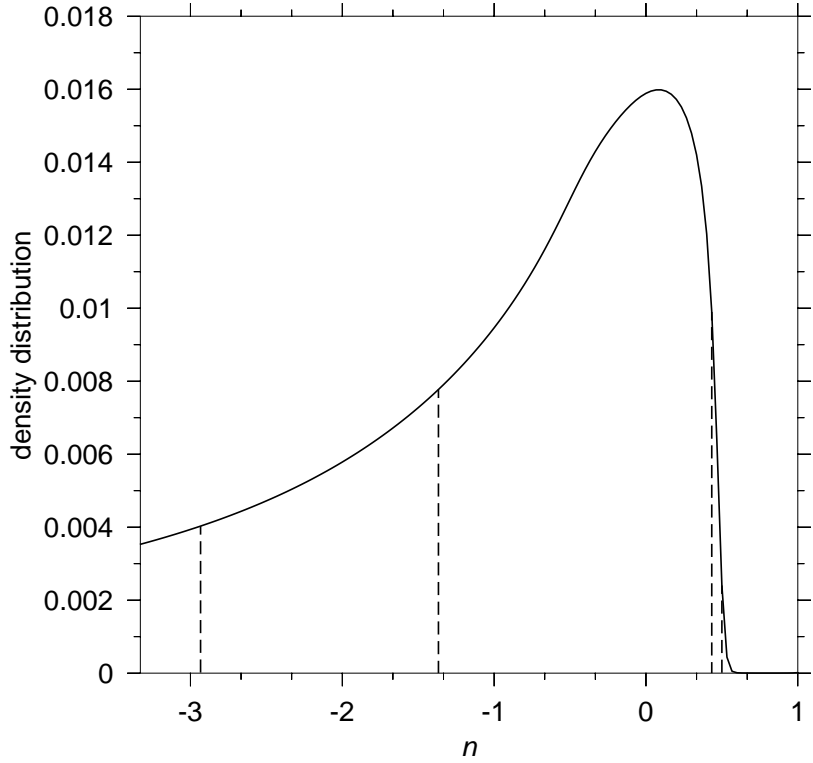}
\caption{The density distribution for $n$ in the Cardassian
model (sample C). We obtain that $n=0.10^{+0.36}_{-1.46}$
on the confidence level $68.3\%$ (the inner dash lines).
Both positive and negative values of $n$ are formally possible.}
\label{fig:3}
\end{figure}

It should be noted that values obtained in both methods are different
but the minimization procedure seem to be more adequate for analyzing our
problem. The density distributions
$f(\Omega_{k,0})$ and $f(n)$ in the Cardassian model are presented on
Fig.~\ref{fig:2} and~\ref{fig:3}, respectively. These figures show that
the preferred model of the Universe is the nearly flat one, that is in
agreement with CMBR data. For sample C the most probable value of
$\Omega_{\mathrm{m},0}$ is $0.29$ that is in agreement with the present
CMBR and extragalactic data \citep{Peebles03,Lahav02}. Therefore, the
detailed analysis of the flat case of the Cardassian model is carried out.

\begin{figure}
\includegraphics[width=0.6\textwidth]{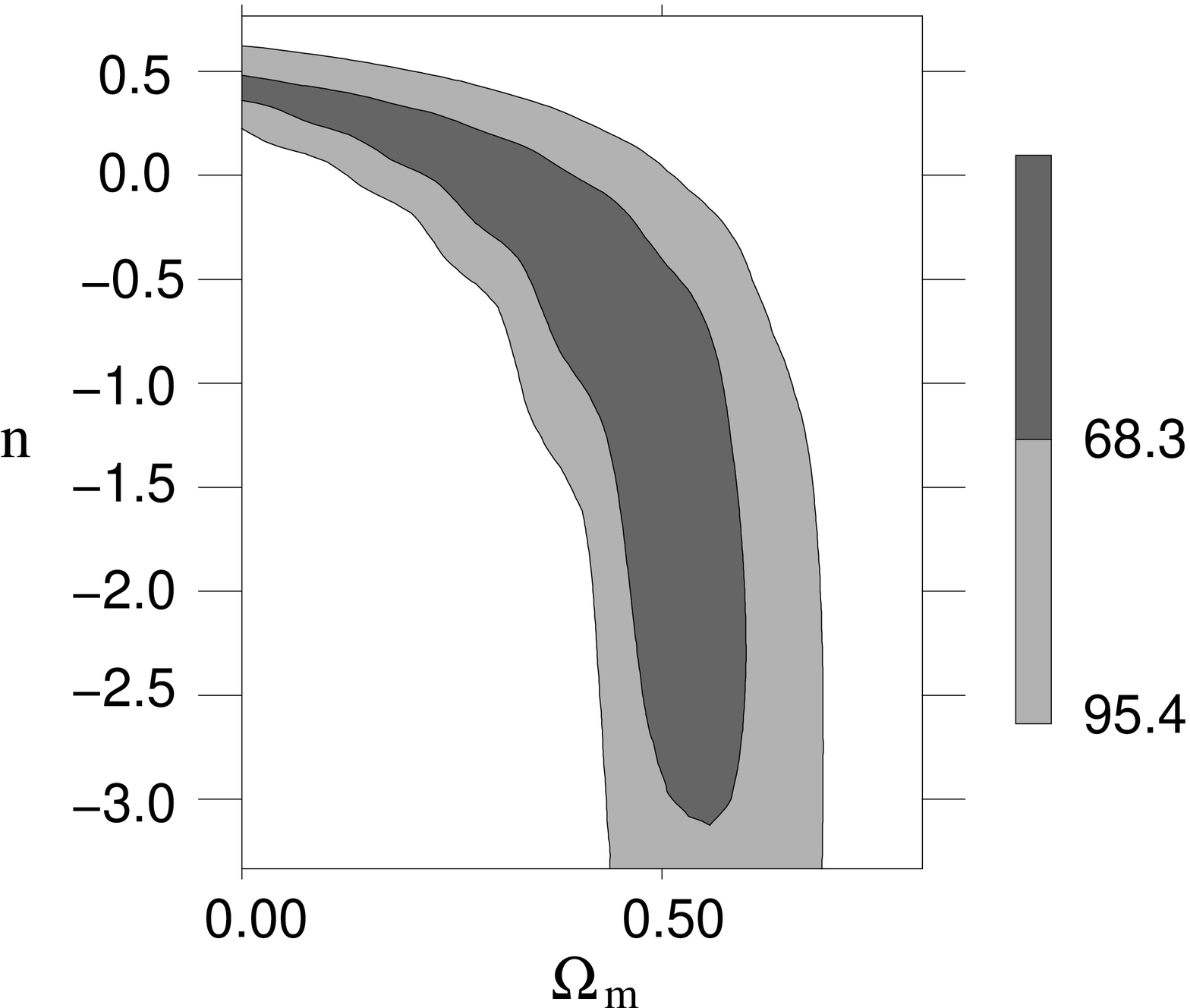}
\caption{Confidence levels on the plane
$(\Omega_{\mathrm{m},0}, n)$ minimized over $\mathcal{M}$
for the flat model ($\Omega_{k,0}=0$), and with
$\Omega_{\mathrm{Card},0}=1-\Omega_{\mathrm{m},0}$.
The figure shows of the preferred value of
$\Omega_{\mathrm{m},0}$ and $n$.}
\label{fig:4}
\end{figure}

In the Fig.~\ref{fig:4} we presented confidence levels on the plane
$(\Omega_{\mathrm{m},0},n)$ minimized over $\mathcal{M}$ for the flat
Cardassian model. Figure~\ref{fig:4} shows the preferred values of parameters
$\Omega_{\mathrm{m},0}$ and $n$. We find that the expected value of
$\Omega_{\mathrm{m},0}$ increases when $n$ decreases. The similar result was
already obtained by \citet{Sen03}, for $n>0$. It should be pointed out
that these authors argued that $\Omega_{\mathrm{m},0}$ could be less than
$0.3$ from the confidence level obtained from the joint analysis of supernovae
data and the positions of the CMBR peaks. In the Cardassian model the
structure formation would be significantly different from, e.g., the classical
model with the cosmological constant. Consequently, in order to calculate
the consequence for the CMB the simple angular rescaling does not suffice.

The detailed results of our analysis for the flat model are summarized in
Table~\ref{tab:2}. The best fit procedure suggests that $n$ should be negative
and consequently $\Omega_{\mathrm{m},0}$ is greater than $0.3$. While we find
the parameter $n$ is negative for minimum value of $\chi^2$ statistic, the
confidence levels allows the positive values of $n$. For the maximum
likelihood method on sample A we obtain that $\Omega_{\mathrm{m},0} =
0.60^{+0.06}_{-0.15}$ and $n=-0.53^{+0.86}_{-1.27}$ on the confidence level
$68.3\%$ for $\mathcal{M}=-3.39$; while $\Omega_{\mathrm{m},0}=
0.54^{+0.10}_{-0.11}$ and $n=-1.20^{+0.53}_{-1.36}$ on the confidence level
$68.3\%$ when we marginalize over $\mathcal{M}$.

\begin{figure}
\includegraphics[width=0.6\textwidth]{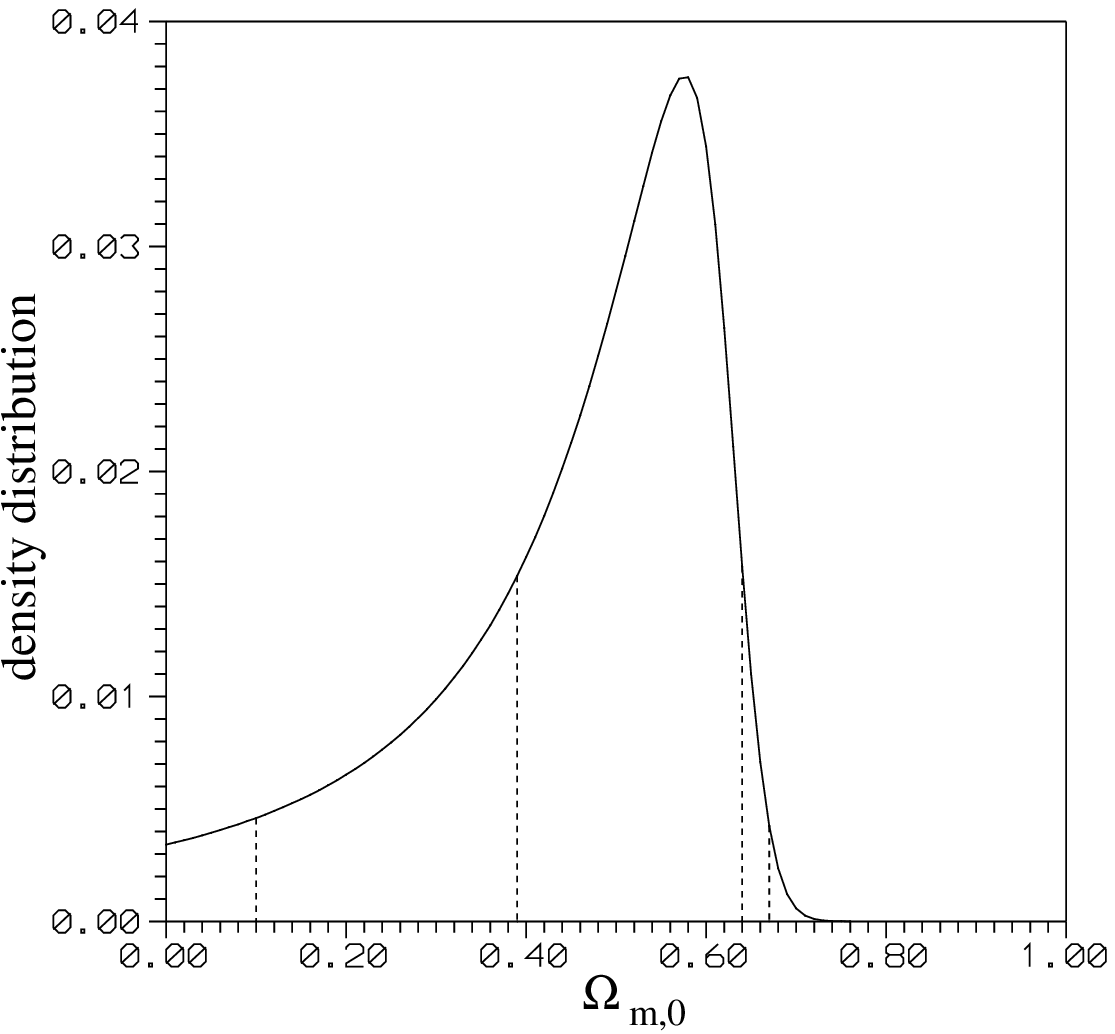}
\caption{The density distribution for $\Omega_{\mathrm{m},0}$ in the Cardassian
flat model. We obtain that $\Omega_{\mathrm{m},0}=0.58^{+0.06}_{-0.13}$
on the confidence level $68.3\%$ (the inner dash lines). Additionally the
confidence level $95.4\%$ is marked (the outer dash lines).}
\label{fig:5}
\end{figure}

\begin{figure}
\includegraphics[width=0.6\textwidth]{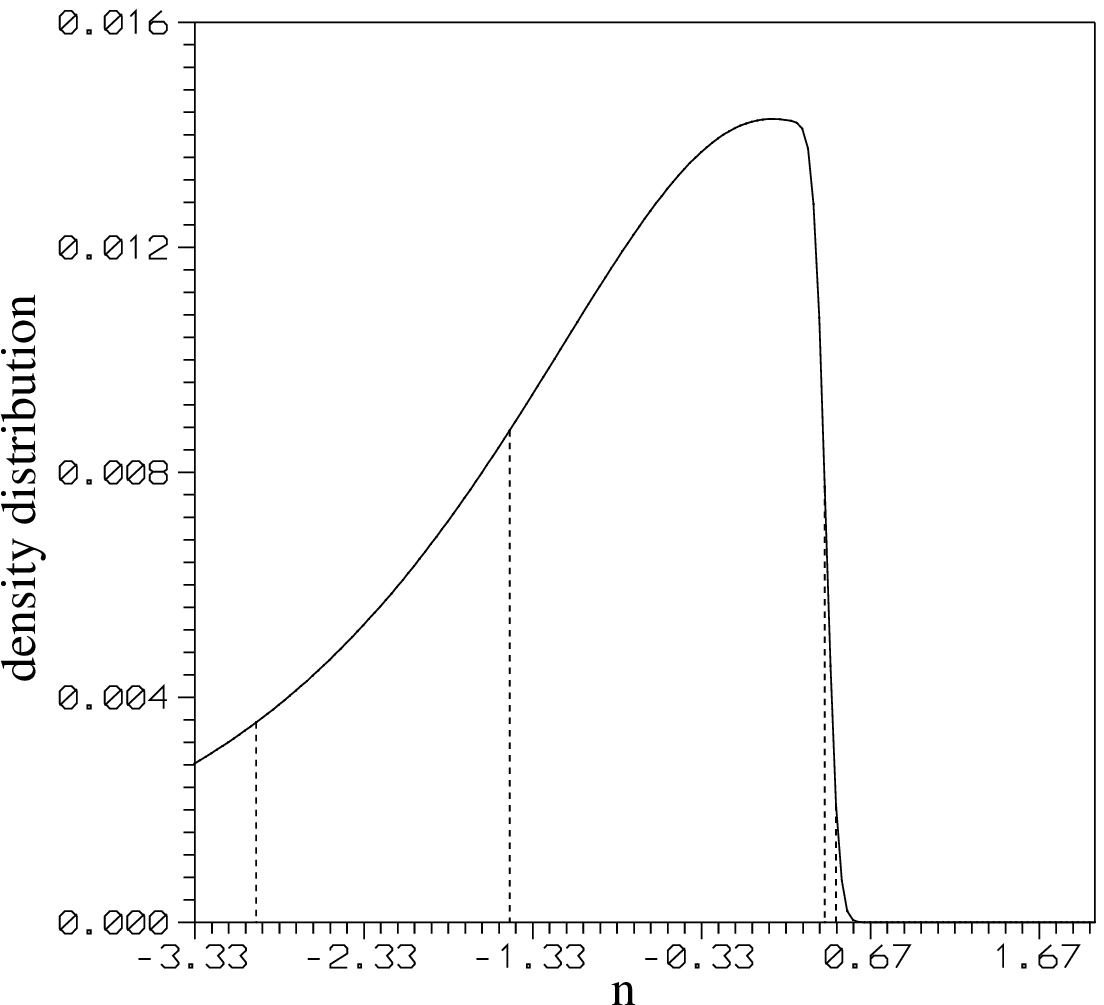}
\caption{The density distribution for $n$ in the Cardassian flat model.
We obtain that $n=0.10^{+0.30}_{-1.57}$ on the confidence level $68.3\%$ (the
inner dash lines). Additionally the confidence level $95.4\%$ is marked (the
outer dash lines). Both positive and negative values of $n$ are formally
possible.}
\label{fig:6}
\end{figure}

In turn for sample C we obtain that $\Omega_{\mathrm{m},0}=
0.58^{+0.06}_{-0.13}$ and $n=0.10^{+0.30}_{-1.57}$ on the confidence level
$68.3\%$ for $\mathcal{M}=-3.43$ (Fig.~\ref{fig:5} and~\ref{fig:6}), while
$\Omega_{\mathrm{m},0}=0.52^{+0.10}_{-0.13}$ and $n=0.17^{+0.20}_{-2.04}$
on the confidence level $68.3\%$ when we marginalize over $\mathcal{M}$.

\begin{figure}
\includegraphics[width=0.6\textwidth]{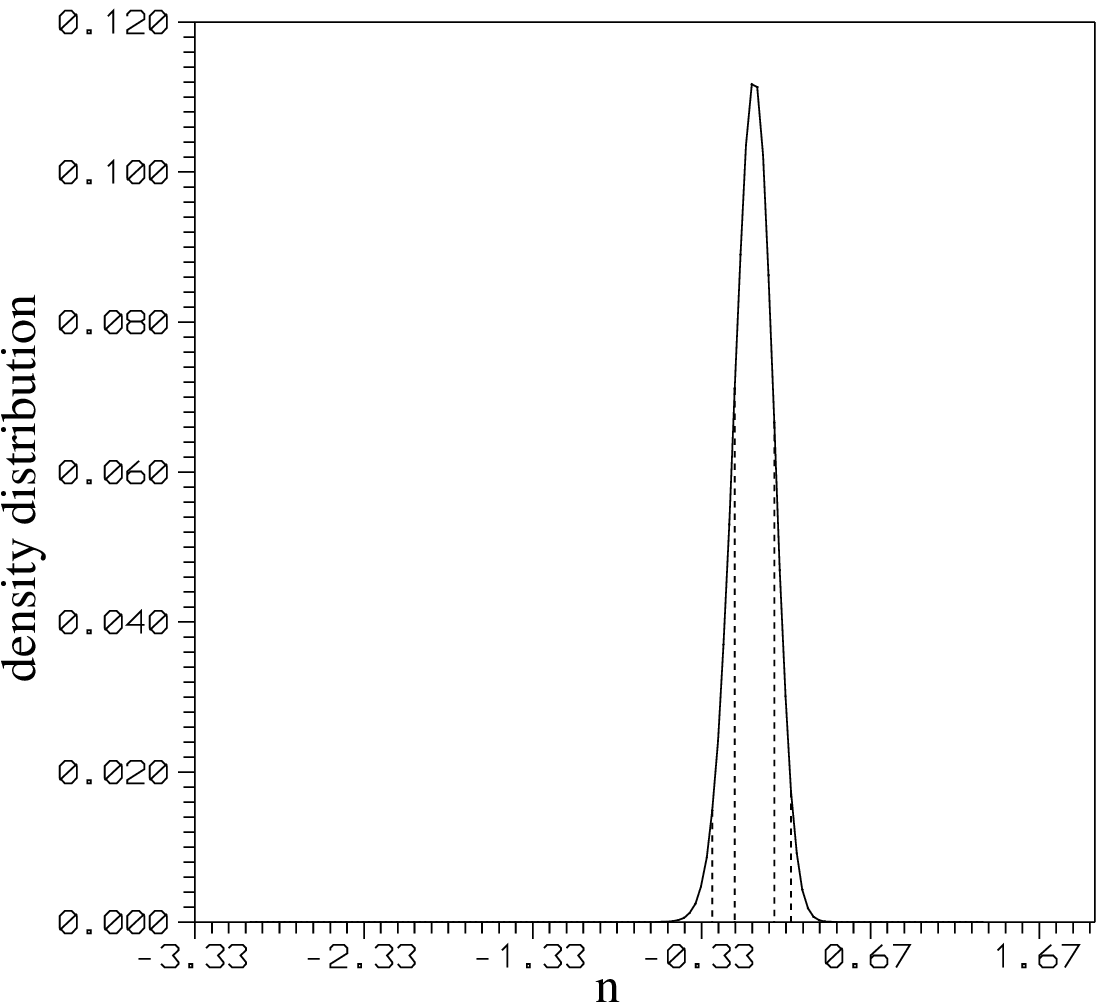}
\caption{The density distribution for $n$ in the Cardassian flat model with
$\Omega_{\mathrm{m},0}=0.3$. We obtain that $n=0.03$ with $\sigma(n)=0.12$.
Both positive and negative values of $n$ are formally possible.}
\label{fig:7}
\end{figure}

For the flat model with $\Omega_{\mathrm{m},0}=0.3$ we obtain for sample A
$n = -0.03$ with $\sigma(n)=0.13$ both for $\mathcal{M}=-3.39$ and
with $\sigma(n)=0.33$ when we marginalize over $\mathcal{M}$.
In turn for sample C we obtain that $n = 0.03$ with $\sigma(n)=0.12$ both
for $\mathcal{M}=-3.43$ (Fig.~\ref{fig:7}) and $\sigma(n)=0.23$ when we
marginalize over $\mathcal{M}$.

\section{Statistical analysis with the Knop sample}

Because the Perlmutter sample was completed four years ago, it would be 
interesting to use newer supernovae observations. Lately \citet{Knop03} 
have reegzamined the Permutter sample with host-galaxy extinction correctly 
applied. They chose from the Perlmutter sample these supernovae which were 
the more securely spectrally identified as type Ia and have reasonable color 
measurements. They also included eleven new high redshift supernovae and 
a well known sample with low redshift supernowae.

We have also decided to test the Cardassian model using this new sample of 
supernovae. They distinguished a few subsets of supernovae from this sample. 
We consider two of them. The first is a subset of 58 supernovae with extinction 
correction (Knop subsample 6; hereafter K6) and the second one of 54 supernovae 
with low extinction (Knop subsample 3; hereafter K3). Sample C and K3 are 
similarly constructed because both contain only low extinction supernovae.

It should be pointed out that in contrast to \citet{Perlmutter99} who included 
errors in measurement of redshift $z$ \citet{Knop03} took only uncertainties in
the redshift due to peculiar velocities. It is the reason that for security we
separately repeated our analysis including errors in measurement of redshift 
$z$ (subsamples K6z and K3z). The errors was taken from \citet{Perlmutter99}. 
In the case when errors were not available we assume $\sigma(z)=0.001$. As one 
can see above, this change has only marginal influence on the results.

At first we estimate the value of $\mathcal{M}$ (equation~(\ref{eq:9})) 
separately for both sample K6 and K3. We obtain the value of $\mathcal{M}=-3.53$
for sample K6, and $\mathcal{M}=-3.48$ for sample K3. It is in agreement
with \citet{Knop03}.

First, we estimate best fitting parameters for non-flat Cardassian model
and obtain $\Omega_{k,0}=-0.24$, $\Omega_{\mathrm{m},0}=0.65$, $n=-1.10$,
$\Omega_{\mathrm{Card},0}=0.59$ for sample K6, and
$\Omega_{k,0}=-0.36$, $\Omega_{\mathrm{m},0}=0.49$, $n=-0.20$,
$\Omega_{\mathrm{Card},0}=0.87$ for sample K3.
When we assume that $\Omega_{k,0}=0$ (flat universe)
then we obtain the best-fitted flat Cardassian model with
$\Omega_{\mathrm{m},0}=0.53$, $n=-1.50$, $\Omega_{\mathrm{Card},0}=0.47$ for
sample K6, and $\Omega_{\mathrm{m},0}=0.41$, $n=-0.63$,
$\Omega_{\mathrm{Card},0}=0.59$ for sample K3. The detailed values are in
Tables~\ref{tab:4} and~\ref{tab:5}.

\begin{table}
\noindent
\caption{Results of the statistical analysis of the Cardassian model
obtained both for the Knop samples from the best fit with minimum
$\chi^2$ (denoted with BF) and from the likelihood method (denoted with L).
The same analysis was repeated with fixed $\Omega_{\mathrm{m},0}=0.3$.}
\label{tab:4}
\begin{tabular}{@{}p{1.5cm}rrrrrrr}
\hline \hline
sample & $\Omega_{\mathrm{Card}}$ & $\Omega_{k,0}$ &
$\Omega_{\mathrm{m},0}$ & $n$ & $\mathcal{M}$ & $\chi^2$& method \\
\hline
 K6   &  0.59 &-0.24 & 0.63 &-1.10 &-3.53 & 54.8 &  BF  \\
      &  0.39 &-0.02 & 0.45 &-0.43 &-3.53 & ---  &  L   \\
      &  0.31 & 0.39 & 0.30 &-2.43 &-3.53 & 54.8 &  BF  \\
      &  0.32 & 0.38 & 0.30 & 0.27 &-3.53 & ---  &  L   \\
\hline
 K6z  &  0.48 &-0.08 & 0.60 &-1.77 &-3.53 & 43.6 &  BF  \\
      &  0.43 &-0.15 & 0.52 &-0.50 &-3.53 & ---  &  L   \\
      &  0.29 & 0.41 & 0.30 &-3.17 &-3.53 & 43.7 &  BF  \\
      &  0.33 & 0.37 & 0.30 & 0.27 &-3.53 & ---  &  L   \\
\hline
 K3   &  0.87 &-0.36 & 0.49 &-0.20 &-3.48 & 60.3 &  BF  \\
      &  0.34 & 0.29 & 0.27 &-0.03 &-3.48 & ---  &  L   \\
      &  0.45 & 0.25 & 0.30 &-0.97 &-3.48 & 60.4 &  BF  \\
      &  0.35 & 0.35 & 0.30 & 0.27 &-3.48 & ---  &  L   \\
\hline
 K3z  &  0.80 &-0.27 & 0.47 &-0.27 &-3.48 & 52.0 &  BF  \\
      &  0.34 & 0.30 & 0.25 & 0.00 &-3.48 & ---  &  L   \\
      &  0.47 & 0.23 & 0.30 &-0.83 &-3.48 & 52.1 &  BF  \\
      &  0.35 & 0.35 & 0.30 & 0.27 &-3.48 & ---  &  L   \\
\end{tabular}
\end{table}

\begin{table}
\noindent
\caption{Results of the statistical analysis of the Cardassian flat model
for the Knop samples obtained both from the best fit with minimum $\chi^2$ 
(denoted with BF) and from the likelihood method (denoted with L).
In the case in which we marginalize over $\mathcal{M}$ we denote it with
$\mathcal{M}$. The same analysis was repeated with fixed
$\Omega_{\mathrm{m},0}=0.3$.}
\label{tab:5}
\begin{tabular}{@{}p{1.5cm}rrrrrr}
\hline \hline
sample & $\Omega_{\mathrm{Card},0}$ & $\Omega_{\mathrm{m},0}$ &
$n$ & $\mathcal{M}$ & $\chi^2$& method \\
\hline
 K6   &  0.47 & 0.53 & -1.50 & -3.53 & 54.6 & BF               \\
      &  0.44 & 0.56 & -1.13 & -3.53 & ---  & L                \\
      &  0.46 & 0.54 & -3.33 & -3.60 & 53.5 & $\mathcal{M}$, BF\\
      &  0.48 & 0.52 & -3.33 & -3.55 & ---  & $\mathcal{M}$, L \\
      &  0.70 & 0.30 & -0.17 & -3.53 & 55.7 & BF               \\
      &  0.70 & 0.30 & -0.17 & -3.53 & ---  & L                \\
      &  0.70 & 0.30 & -0.10 & -3.52 & 55.6 & $\mathcal{M}$, BF\\
      &  0.70 & 0.30 & -0.13 & -3.53 & ---  & $\mathcal{M}$, L \\
\hline
 K6z  &  0.45 & 0.55 & -1.87 & -3.53 & 43.6 & BF               \\
      &  0.43 & 0.57 & -1.57 & -3.53 & ---  & L                \\
      &  0.46 & 0.54 & -3.33 & -3.60 & 42.7 & $\mathcal{M}$, BF\\
      &  0.48 & 0.52 & -3.33 & -3.54 & ---  & $\mathcal{M}$, L \\
      &  0.70 & 0.30 & -0.17 & -3.53 & 44.9 & BF               \\
      &  0.70 & 0.30 & -0.17 & -3.53 & ---  & L                \\
      &  0.70 & 0.30 & -0.07 & -3.51 & 44.8 & $\mathcal{M}$, BF\\
      &  0.70 & 0.30 & -0.10 & -3.51 & ---  & $\mathcal{M}$, L \\
\hline
 K3   &  0.59 & 0.41 & -0.63 & -3.48 & 60.3 & BF               \\
      &  0.50 & 0.50 & -0.37 & -3.48 & ---  & L                \\
      &  0.58 & 0.42 & -0.77 & -3.49 & 60.3 & $\mathcal{M}$, BF\\
      &  0.52 & 0.48 & -0.40 & -3.49 & ---  & $\mathcal{M}$, L \\
      &  0.70 & 0.30 & -0.20 & -3.48 & 60.4 & BF               \\
      &  0.70 & 0.30 & -0.20 & -3.48 & ---  & L                \\
      &  0.70 & 0.30 & -0.13 & -3.47 & 60.4 & $\mathcal{M}$, BF\\
      &  0.70 & 0.30 & -0.17 & -3.48 & ---  & $\mathcal{M}$, L \\
\hline
 K3z  &  0.60 & 0.40 & -0.57 & -3.48 & 52.0 & BF               \\
      &  0.49 & 0.51 & -0.30 & -3.48 & ---  & L                \\
      &  0.58 & 0.42 & -0.83 & -3.50 & 52.0 & $\mathcal{M}$, BF\\
      &  0.52 & 0.48 & -0.40 & -3.50 & ---  & $\mathcal{M}$, L \\
      &  0.70 & 0.30 & -0.17 & -3.48 & 52.2 & BF               \\
      &  0.70 & 0.30 & -0.17 & -3.48 & ---  & L                \\
      &  0.70 & 0.30 & -0.17 & -3.48 & 52.2 & $\mathcal{M}$, BF\\
      &  0.70 & 0.30 & -0.17 & -3.48 & ---  & $\mathcal{M}$, L \\
\end{tabular}
\end{table}

We also carry out the model parameters estimation using the minimization
procedure, based on the likelihood method. On the confidence level $68.3\%$ 
we obtain parameter values for samples K6 and K3 (Table~\ref{tab:6}).

\begin{table}
\caption{Model parameter values obtained from the minimization procedure
carried out on the Knop samples.}
\label{tab:6}
\begin{tabular}{@{}p{1.5cm}rrrr}
\hline \hline
sample & $\Omega_{k,0}$ & $\Omega_{\mathrm{Card},0}$ &
$\Omega_{\mathrm{m},0}$ & $n$ \\
\hline
 K6 &  $-0.02^{+0.49}_{-0.54}$ & $0.39^{+0.34}_{-0.18}$
    & $0.45^{+0.30}_{-0.31}$ & $-0.43^{+0.73}_{-1.50}$ \\
 K6z&  $-0.15^{+0.56}_{-0.45}$ & $0.43^{+0.28}_{-0.21}$
    & $0.52^{+0.30}_{-0.32}$ & $-0.50^{+0.63}_{-1.63}$ \\
 K3 & $0.29^{+0.38}_{-0.66}$ & $0.34^{+0.44}_{-0.13}$
    & $0.27^{+0.13}_{-0.23}$ & $-0.03^{+0.33}_{-1.37}$ \\
 K3z& $0.30^{+0.38}_{-0.66}$ & $0.34^{+0.45}_{-0.14}$
    & $0.25^{+0.23}_{-0.23}$ & $-0.00^{+0.40}_{-1.40}$
\end{tabular}
\end{table}

It should be noted that values obtained in both methods are different
but differences are much smaller then in the case of original Perlmutter
sample.

The detailed results of our analysis for the flat model are summarized in
Table~\ref{tab:5}. The best fit procedure suggests that $n$ should be negative
and consequently $\Omega_{\mathrm{m},0}$ is greater than $0.3$. While we find
the parameter $n$ is negative for minimum value of $\chi^2$ statistic, the
confidence levels allows the positive values of $n$. For the maximum
likelihood method on sample K6 we obtain that $\Omega_{\mathrm{m},0} =
0.44^{+0.14}_{-0.08}$ and $n=-1.13^{+1.07}_{-1.13}$ on the confidence level
$68.3\%$ for $\mathcal{M}=-3.53$; while $\Omega_{\mathrm{m},0}=
0.48^{+0.09}_{-0.09}$ and $n=-3.33^{+2.00}$ on the confidence level
$68.3\%$ when we marginalize over $\mathcal{M}$.

In turn for sample K3 we obtain that $\Omega_{\mathrm{m},0}=
0.50^{+0.07}_{-0.15}$ and $n=-0.37^{+0.63}_{-1.00}$ on the confidence level
$68.3\%$ for $\mathcal{M}=-3.48$, while
$\Omega_{\mathrm{m},0}=0.48^{+0.08}_{-0.13}$ and $n=-0.40^{+0.77}_{-1.24}$
on the confidence level $68.3\%$ when we marginalize over $\mathcal{M}$.

For the flat model with $\Omega_{\mathrm{m},0}=0.3$ we obtain for sample K6
$n = -0.17$ with $\sigma(n)=0.16$ for $\mathcal{M}=-3.39$ and $n = -0.13$
with $\sigma(n)=0.23$ when we marginalize over $\mathcal{M}$.
In turn for sample K3 we obtain that $n =-0.20$ with $\sigma(n)=0.13$
for $\mathcal{M}=-3.48$ and $n =-0.20$ $\sigma(n)=0.17$ when we
marginalize over $\mathcal{M}$.

\section{Statistical analysis with the Tonry sample}
 
Another sample was presented by \citet{Tonry03} who 
collected a large number of supernovae published by different authors 
and added eight new high redshift SN Ia. This sample of 230 SNe Ia 
was recalibrated with consistent zero point. Whenever it was possible 
the extinctions estimates and distance fitting were recomputed. However, none 
of the methods was able to apply to all supernovae (for details see 
Tab.~8 \cite{Tonry03}).
 
Despite of this problem, the analysis of the Cardassian model using
this sample of supernovae could be interesting. We decide to analyse 
four subsamples. First, the full Tonry sample of 230 SNe Ia
(hereafter sample Ta) is considered. The sample of 197 SNe Ia (hereafter
sample Tb) consists of low extinction supernovae only (median $V$ band 
extinction $A_V<0.5$). 
Because the Tonry sample has a lot of outliers especially in low redshift, 
the sample of 195 SN Ia is such that all low redshift ($z<0.01$) supernovae 
are excluded (hereafter sample Tc). 
In the sample of 172 SN Ia all supernovae with low redshift and and high 
extinction are omitted (hereafter sample Td).

\citet{Tonry03} presented redshift and luminosity distance observations 
for their sample of supernovae. Therefore, equations 
(\ref{eq:8}) and (\ref{eq:9}) should be modified \cite{Williams03}:
\begin{equation}
\label{eq:13a}
m-M = 5\log_{10}(\mathcal{D}_L)_{\mathrm{Tonry}}-5\log_{10}65 + 25
\end{equation}
and 
\begin{equation}
\label{eq:13b}
\mathcal{M}=-5\log_{10}H_0+25.
\end{equation}
For $H_0=65$ km s$^{-1}$ Mpc$^{-1}$ we obtain $\mathcal{M}=15.935$.

First, we estimate best fitting parameters for the non-flat Cardassian model
and obtain $\Omega_{k,0}=-1$, $\Omega_{\mathrm{m},0}=0.48$, $n=-0.27$,
$\Omega_{\mathrm{Card},0}=1.52$ for sample Ta.
When we assume that $\Omega_{k,0}=0$ (flat universe)
then for the sample Ta we obtain the best-fitted flat Cardassian model with
$\Omega_{\mathrm{m},0}=0.46$, $n=-1.50$, $\Omega_{\mathrm{Card},0}=0.54$.
For samples Tb, Tc, and Td results differs only marginally.
The detailed values are presented in Tables~\ref{tab:7} and~\ref{tab:8}.

\begin{table}
\noindent
\caption{Results of the statistical analysis of the Cardassian model
obtained for the Tonry samples both from the best fit with minimum
$\chi^2$ (denoted with BF) and from the likelihood method (denoted with L).
The same analysis was repeated with fixed $\Omega_{\mathrm{m},0}=0.3$.}
\label{tab:7}
\begin{tabular}{@{}p{1.5cm}rrrrrrr}
\hline \hline
sample & $\Omega_{\mathrm{Card}}$ & $\Omega_{k,0}$ &
$\Omega_{\mathrm{m},0}$ & $n$ & $\mathcal{M}$ & $\chi^2$& method \\
\hline
 Ta   &  1.52 &-1.00 & 0.48 & 0.27 &15.935&252.1 &  BF  \\
      &  0.54 &-0.64 & 0.49 & 0.10 &15.935& ---  &  L   \\
      &  1.68 &-0.98 & 0.30 & 0.37 &15.935&252.5 &  BF  \\
      &  0.28 & 0.42 & 0.30 & 0.33 &15.935& ---  &  L   \\
\hline
 Tb   &  1.50 &-1.00 & 0.50 & 0.23 &15.935&185.3 &  BF  \\
      &  0.64 &-0.78 & 0.49 & 0.07 &15.935& ---  &  L   \\
      &  1.70 &-1.00 & 0.30 & 0.37 &15.935&185.7 &  BF  \\
      &  0.31 & 0.39 & 0.30 & 0.33 &15.935& ---  &  L   \\
\hline
 Tc   &  1.47 &-1.00 & 0.48 & 0.23 &15.935&201.0 &  BF  \\
      &  0.54 &-0.64 & 0.49 & 0.10 &15.935& ---  &  L   \\
      &  1.68 &-0.98 & 0.53 & 0.37 &15.935&201.3 &  BF  \\
      &  0.28 & 0.42 & 0.49 & 0.33 &15.935& ---  &  L   \\
\hline
 Td   &  1.50 &-1.00 & 0.50 & 0.23 &15.935&164.9 &  BF  \\
      &  0.63 &-0.78 & 0.49 & 0.17 &15.935& ---  &  L   \\
      &  1.70 &-1.00 & 0.30 & 0.37 &15.935&165.4 &  BF  \\
      &  0.31 & 0.39 & 0.30 & 0.33 &15.935& ---  &  L   \\
\end{tabular}
\end{table}
 
\begin{table}
\noindent
\caption{Results of the statistical analysis of the Cardassian flat model
for the Tonry samples obtained both from the best fit with minimum $\chi^2$ 
(denoted with BF) and from the likelihood method (denoted with L).
In the case in which we marginalize over $\mathcal{M}$ we denote it with
$\mathcal{M}$. The same analysis was repeated with fixed
$\Omega_{\mathrm{m},0}=0.3$.}
\label{tab:8}
\begin{tabular}{@{}p{1.5cm}rrrrrr}
\hline \hline
sample & $\Omega_{\mathrm{Card},0}$ & $\Omega_{\mathrm{m},0}$ &
$n$ & $\mathcal{M}$ & $\chi^2$& method \\
\hline
 Ta   &  0.54 & 0.46 & -0.50 & 15.935&253.6 & BF               \\
      &  0.49 & 0.51 & -0.40 & 15.935& ---  & L                \\
      &  0.48 & 0.52 & -2.00 & 15.855&247.7 & $\mathcal{M}$, BF\\
      &  0.47 & 0.53 & -1.87 & 15.855& ---  & $\mathcal{M}$, L \\
      &  0.70 & 0.30 &  0.03 & 15.935&254.7 & BF               \\
      &  0.70 & 0.30 &  0.03 & 15.935& ---  & L                \\
      &  0.70 & 0.30 & -0.10 & 15.895&252.1 & $\mathcal{M}$, BF\\
      &  0.70 & 0.30 & -0.10 & 15.895& ---  & $\mathcal{M}$, L \\
\hline
 Tb   &  0.56 & 0.44 & -0.50 & 15.935&187.1 & BF               \\
      &  0.51 & 0.49 & -0.40 & 15.935& ---  & L                \\
      &  0.51 & 0.49 & -1.47 & 15.875&183.4 & $\mathcal{M}$, BF\\
      &  0.49 & 0.51 & -1.40 & 15.875& ---  & $\mathcal{M}$, L \\
      &  0.70 & 0.30 & -0.03 & 15.935&188.1 & BF               \\
      &  0.70 & 0.30 & -0.03 & 15.935& ---  & L                \\
      &  0.70 & 0.30 & -0.13 & 15.905&186.5 & $\mathcal{M}$, BF\\
      &  0.70 & 0.30 & -0.17 & 15.905& ---  & $\mathcal{M}$, L \\
\hline
 Tc   &  0.54 & 0.46 & -0.50 & 15.935&202.5 & BF               \\
      &  0.49 & 0.51 & -0.40 & 15.935& ---  & L                \\
      &  0.49 & 0.51 & -1.53 & 15.875&198.7 & $\mathcal{M}$, BF\\
      &  0.47 & 0.53 & -1.47 & 15.875& ---  & $\mathcal{M}$, L \\
      &  0.70 & 0.30 &  0.03 & 15.935&203.6 & BF               \\
      &  0.70 & 0.30 &  0.03 & 15.935& ---  & L                \\
      &  0.70 & 0.30 & -0.07 & 15.905&202.2 & $\mathcal{M}$, BF\\
      &  0.70 & 0.30 & -0.07 & 15.905& ---  & $\mathcal{M}$, L \\
\hline
 Td   &  0.56 & 0.44 & -0.50 & 15.935&166.7 & BF               \\
      &  0.51 & 0.49 & -0.43 & 15.935& ---  & L                \\
      &  0.52 & 0.48 & -1.23 & 15.885&164.1 & $\mathcal{M}$, BF\\
      &  0.49 & 0.51 & -1.20 & 15.885& ---  & $\mathcal{M}$, L \\
      &  0.70 & 0.30 & -0.03 & 15.935&167.7 & BF               \\
      &  0.70 & 0.30 & -0.03 & 15.935& ---  & L                \\
      &  0.70 & 0.30 & -0.13 & 15.905&166.8 & $\mathcal{M}$, BF\\
      &  0.70 & 0.30 & -0.13 & 15.905& ---  & $\mathcal{M}$, L \\
\end{tabular}
\end{table}
 
As in previous sections,
we also carry out the model parameters estimation using the minimization
procedure based on the likelihood method. On the confidence level $68.3\%$ we
presented obtained parameter values for Tonry samples in the Table~\ref{tab:9}.
 
\begin{table}
\caption{Model parameter values obtained from the minimization procedure
carried out on the Tonry sample. To obtain errors of the parameter 
$\Omega_{k,0}$ we enlarge an estimation interval of $\Omega_{k,0}$ to 
$[-2, 2]$.}
\label{tab:9}
\begin{tabular}{@{}p{1.5cm}rrrr}
\hline \hline
sample & $\Omega_{k,0}$ & $\Omega_{\mathrm{Card},0}$ &
$\Omega_{\mathrm{m},0}$ & $n$ \\
\hline
 Ta & $-0.64^{+0.72}_{-0.61}$ & $0.54^{+0.52}_{-0.27}$
    & $ 0.49^{+0.21}_{-0.23}$ & $0.10^{+0.36}_{-0.70}$ \\
 Tb & $-0.78^{+0.76}_{-0.55}$ & $0.64^{+0.50}_{-0.32}$
    & $ 0.49^{+0.21}_{-0.23}$ & $0.07^{+0.36}_{-0.67}$ \\
 Tc & $-0.64^{+0.72}_{-0.61}$ & $0.54^{+0.52}_{-0.27}$
    & $ 0.49^{+0.21}_{-0.22}$ & $0.10^{+0.33}_{-0.70}$ \\
 Td & $-0.78^{+0.77}_{-0.55}$ & $0.63^{+0.51}_{-0.31}$
    & $ 0.49^{+0.21}_{-0.23}$ & $0.17^{+0.28}_{-0.87}$
\end{tabular}
\end{table}

The detailed results of our analysis for the flat model, based on the 
likelihood method, are summarized in Table~\ref{tab:8}. For sample Ta 
we obtain that $\Omega_{\mathrm{m},0} =
0.51^{+0.08}_{-0.12}$ and $n=-0.40^{+0.80}_{-0.63}$ on the confidence level
$68.3\%$ for $\mathcal{M}=15.935$; while $\Omega_{\mathrm{m},0}=
0.53^{+0.04}_{-0.06}$ and $n=-1.87^{+0.83}_{-1.00}$ on the confidence level
$68.3\%$ when we marginalize over $\mathcal{M}$.

In turn for sample Td we obtain that $\Omega_{\mathrm{m},0}=
0.49^{+0.08}_{-0.12}$ and $n=-0.43^{+0.50}_{-0.63}$ on the confidence level
$68.3\%$ for $\mathcal{M}=15.935$, while
$\Omega_{\mathrm{m},0}=0.51^{+0.05}_{-0.06}$ and $n=-1.20^{+0.77}_{-1.06}$
on the confidence level $68.3\%$ when we marginalize over $\mathcal{M}$.
 
For the flat model with $\Omega_{\mathrm{m},0}=0.3$ we obtain for subsample 
Ta $n = 0.03$ with $\sigma(n)=0.10$ for $\mathcal{M}=15.935$ and $n = -0.10$
with $\sigma(n)=0.15$ when we marginalize over $\mathcal{M}$.
For other subsamples results are very similar.

All these results suggest that $n$ is negative and consequently 
$\Omega_{\mathrm{m},0}$ is greater than $0.3$. 

Generally we find that results obtained with the Tonry sample are the
similar as obtained with the Perlmutter sample apart from the non flat 
case where values of $n$ parameter are positive, because 
either positive and negative values are in the Perlmutter samples.

\section{Discussion}

\begin{figure}
\includegraphics[width=\textwidth]{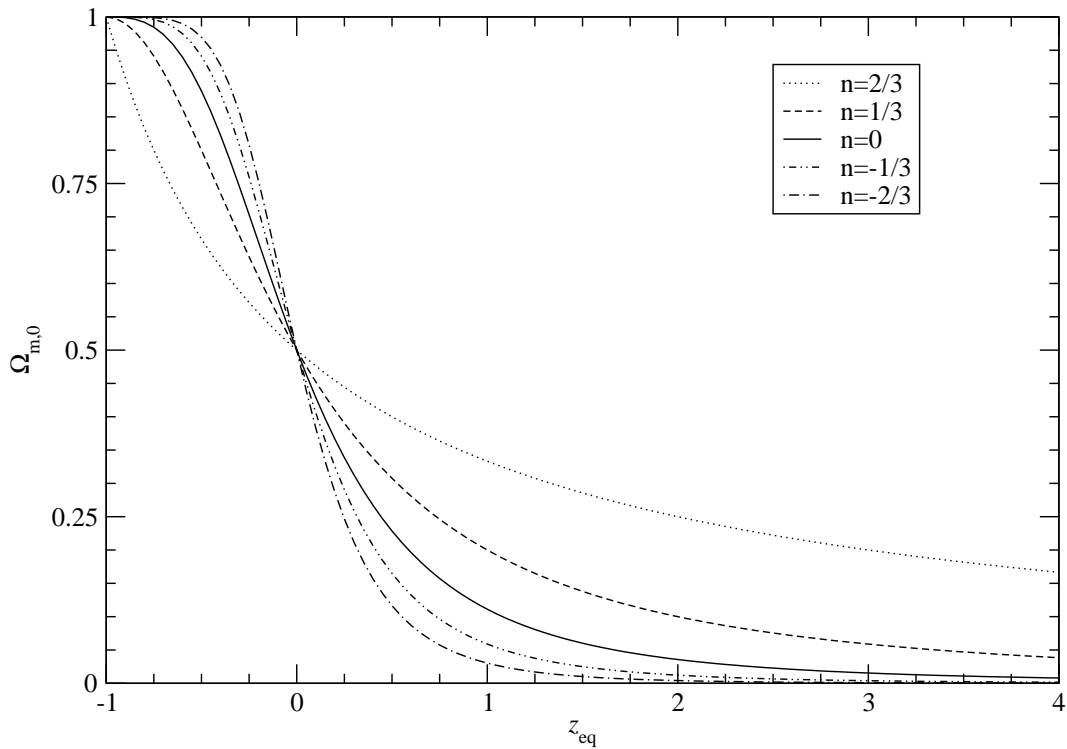}
\caption{Relation between $\Omega_{\mathrm{m},0}$ and $z_{\mathrm{eq}}$ for
different $n$.}
\label{fig:8}
\end{figure}

In the present paper we discussed the problem of universe acceleration
in the Cardassian model. Our results are different from those obtained by
\citet{Zhu03} who suggested the universe with very low matter density.
Our results indicate the high or normal ($\Omega_{\mathrm{m},0} \approx 0.3$) 
matter density. This difference may come from not including the errors in 
redshifts and using a different variable by them. 
In our fitting procedure we use the more natural variable
$\Omega_{\mathrm{Card},0}$ (or $\Omega_{\mathrm{m},0}$) instead of 
their parameter $z_{\mathrm{eq}}$
($z_{\mathrm{eq}} \colon \frac{\rho}{3}=B\rho^n, \rho=\rho_0 a^{-3},
1+z=a^{-1}$). Two parameters $\Omega_{\mathrm{m},0}$ and
$z_{\mathrm{eq}}$ are related by the formula
\begin{equation}
\label{eq:14}
\Omega_{\mathrm{m},0}=(1+(1+z_{\mathrm{eq}})^{3(1-n)})^{-1}.
\end{equation}
In Fig.~\ref{fig:8} relation~(\ref{eq:14}) for different $n$ is
presented. We can observe that taking a constant step in $z_{\mathrm{eq}}$
it leads to varying steps in $\Omega_{\mathrm{m},0}$. In our method we have
\begin{equation}
\label{eq:15}
\Omega_{\mathrm{m},0} = 1 - \Omega_{\mathrm{Card},0} - \Omega_{k,0}
\end{equation}
and we take the constant step of $0.01$ in $\Omega_{\mathrm{m},0}$. 
For the flat Cardassian model both \citet{Zhu03} and we obtained negative 
values of $n$ with the best fit method. 

Assuming the curvature term $\Omega_{k,0}$ in the Cardassian model we 
obtained its statistical estimation to be close to zero. In this natural way
we got that a nearly flat universe is preferred. Moreover, in this case 
for Perlmutter sample C the preferred value of $\Omega_{\mathrm{m},0}$ is 
$0.29$ that is in agreement with CMBR and extragalactic data. While for 
sample A we got that a universe with higher matter content is preferred.

Let us note that \citet{Riess98,Perlmutter99} obtained the high negative
value of $\Omega_{k,0}$ as the best fit, although zero value of 
$\Omega_{k,0}$ is statistically admissible. To find the curvature they 
additionally used the data from CMBR and extragalactic astronomy. But in 
the Cardassian model in natural way the nearly flat universe is favored.

If the flat universe is assumed, in the analysis of the Cardassian
universe with sample C we got low but positive values of $n$ and matter 
density higher than $0.3$. For all other samples we obtained negative 
values of $n$ (it would also indicate the phamtom fluid with super negative 
pressure). In turn, when $\Omega_{\mathrm{m},0}$ is fixed to be 
$0.3$, then $n \approx 0$ is favored.

In general, the results obtained using the Perlmutter and Knop samples are 
not significantly different. The main advantage of using the Knop sample 
is the lower errors of estimated parameters in the model. 
The Tonry sample also gave the results similar to the Perlmutter sample. 
It can be noted that the Tonry sample does not improve the estimation 
errors compared with the Perlmutter sample. 

From the statistical analysis we found that the Perlmutter and Cardassian 
models are indistinguishable. Hence, using the philosophy of Occam's razor the
Cardassian model should be rejected if we base on current SN Ia data. However,
the Cardassian model has its own advantages and fits well to the present
observational data. And till we have no more precise data we should wait
before we reject this model.

\end{document}